%% file: main.tex
\documentclass[a4paper]{article}
\usepackage[utf8]{inputenc}

  \usepackage[pdftex]{graphicx}
\usepackage[cmex10]{amsmath}
\usepackage[sort,nocompress]{cite}
\usepackage{amssymb}
\usepackage{multirow}
\usepackage[caption=false,font=footnotesize]{subfig}
\usepackage{stfloats}
\usepackage{hyperref}
\usepackage{framed}
\usepackage{color}
\usepackage[dvipsnames]{xcolor}
\usepackage{xspace}

\usepackage{makecell}
\usepackage{multirow}
\usepackage{fullpage}

\newcommand{\arch}{SAFE\xspace}
\newcommand{\archext}{SAFE: Self-Attentive Function Embeddings\xspace}

\newcommand{\normalDataset}{\textsf{AMD64ARMOpenSSL Dataset}\xspace}
\newcommand{\bigDataset}{\textsf{AMD64multipleCompilers Dataset}\xspace}
\newcommand{\postgres}{\textsf{AMD64PostgreSQL Dataset}\xspace}
\newcommand{\simDataset}{\textsf{Semantic Dataset}\xspace}

\newif\ifshowcomments
\showcommentstrue

\ifshowcomments
\newcommand{\mynote}[2]{\fbox{\bfseries\sffamily\scriptsize{#1}}
 {\small$\blacktriangleright$\textsf{\emph{#2}}$\blacktriangleleft$}}
\else
\newcommand{\mynote}[2]{}
\fi

\begin{document}
%
\title{ \archext for Binary Similarity}

\author{ Luca Massarelli$^{\dagger}$, Giuseppe Antonio Di Luna$^{\ddagger}$, Fabio Petroni$^{*}$,\\ Leonardo Querzoni$^\dagger$, Roberto Baldoni$^{\dagger}$\\
\small${\dagger}$: University of Rome Sapienza. \texttt{\{massarelli, querzoni, baldoni\}@diag.uniroma1.it}.\\
\small${\ddagger}$: CINI, National Laboratory of Cyber Security. \texttt{g.a.diluna@gmail.com}.\\
\small$*$: Facebook AI Research, \texttt{petronif@acm.org}.}
\date{}
\maketitle

\begin{abstract}



The binary similarity problem consists in determining if two functions are similar  by only considering their compiled form. Advanced techniques for binary similarity recently gained momentum as they can be applied in several fields, such as copyright disputes, malware analysis, vulnerability detection, etc., and thus have an immediate practical impact. Current solutions compare functions by first transforming their binary code in multi-dimensional vector representations (embeddings), and then comparing vectors through simple and efficient geometric operations. However, embeddings are usually derived from binary code using manual feature extraction, that may fail in considering important function characteristics, or may consider features that are not important for the binary similarity problem. In this paper we propose \arch, a novel architecture for the embedding of functions based on a self-attentive neural network. \arch works directly on disassembled binary functions, does not require manual feature extraction, is computationally more efficient than existing solutions (i.e., it does not incur in the computational overhead of building or manipulating control flow graphs), and is more general as it works on stripped binaries and on multiple architectures. 
We report the results from a quantitative and qualitative analysis that show how \arch provides a noticeable performance improvement with respect to previous solutions.
Furthermore, we show how clusters of our embedding vectors are closely related to the semantic of the implemented algorithms, paving the way for further interesting applications (e.g. semantic-based binary function search)

\end{abstract}


%


\input{sec1_introduction}
\input{sec2_relatedwork}
\input{sec3_similaritydefinition}
\input{sec4_network}
\input{sec5_evaluation.tex}

 \input{sec6_conclusion.tex}

\bibliographystyle{IEEEtranS}
 \bibliography{mybibfile}  

\end{document}

%% file: sec1_introduction.tex
\section{Introduction}
In the last years there has been an exponential increase in the creation of new contents. As all products, also software is subject to this trend. As an example, the number of apps available on the Google Play Store increased from 30K in 2010 to 3 millions in 2018\footnote{\url{https://www.statista.com/statistics/266210/number-of-available-applications-in-the-google-play-store/}}.
This increase directly leads to more vulnerabilities as reported by CVE\footnote{\phantom{}\url{https://www.cvedetails.com/browse-by-date.php}} that witnessed a 120\% growth in the number of discovered vulnerabilities from 2016 to 2017. At the same time complex software spreads in several new devices: the \emph{Internet of Things} has multiplied the number of architectures on which the same program has to run and COTS software components are increasingly integrated in closed-source products.

This multidimensional increase in quantity, complexity and diffusion of software makes the resulting infrastructures difficult to manage and control, as part of their internals are often inaccessible for inspection to their own administrators.
As a consequence, system integrators are looking forward to novel solutions that take into account such issues and provide functionalities to automatically analyze software artifacts in their compiled form (binary code). One prototypical problem in this regard, is the one of {\em binary similarity} \cite{bindiff,khoo,sigma}, where the goal is to find similar functions in compiled code fragments. 

Binary similarity has been recently subject to a lot of attention \cite{David,David3,blanket}. This is due to its centrality in several tasks, such as discovery of known vulnerabilities in large collection of software, dispute on copyright matters, analysis and detection of malicious software, etc.

In this paper, in accordance with \cite{genius} and \cite{CCS}, we focus on a specific version of the binary similarity problem in which we define two binary functions to be similar if they are compiled  from the same source code. As already pointed out in \cite{CCS}, this assumption does not make the problem trivial.

Inspired by \cite{genius} we look for solutions that solve the binary similarity problem using {\em embeddings}.  Loosely speaking, each binary function is first transformed into a vector of numbers (an \emph{embedding}), in such a way that code compiled from a same source results in vectors that are similar.

This idea has several advantages. First of all, once the embeddings are computed, checking the similarity is relatively cheap and fast (we consider the scalar product of two constants size vectors as a constant time operation). Thus we can pre-compute a large collection of embeddings for interesting functions and check against such collection in linear time. In the light of the many use cases, this characteristic is extremely useful.
Another advantage comes from the fact that such embeddings can be used as input to other machine learning algorithms, that can in turn cluster functions, classify them, etc.

Current solutions that adopt this approach, still come with several shortcomings:

\begin{itemize}
	\item they \cite{CCS} use manually selected features to calculate the embeddings, introducing potential bias in the resulting vectors. Such bias stems form the possibility of overlooking important features (that don't get selected), or including features that are expensive to process while not providing noticeable performance improvements; for example, including features extracted from the function's {\em control flow graph} (CFG) imposes a $~10\times$ speed penalty with respect to features extracted from the disassembled code\footnote{Tests conducted using the Radare2 disassembler\cite{radare2}.};
	\item they \cite{asm2vec} assume that call symbols to dynamically linking libraries are available in binary functions (such as libc,msvc,ecc..), while this is not true for binaries that are stripped and statically linked\footnote{Conversely, recognizing library functions in stripped statically linked binaries is an application of the binary similarity problem without symbolic calls.} or in partial fragments of binaries (e.g. extracted during volatile memory forensic analyses);
	\item they work only on specific CPU architectures \cite{asm2vec}.
\end{itemize}




Considering these shortcomings, in this paper we introduce \archext, a solution we designed to overcome all of them. In particular we considered two specific goals:
\begin{itemize}
\item design a solution to quickly generate embeddings for several hundreds of binaries;
\item design a solution that could be applicable the vast majority of cases, i.e. able to work with  stripped binaries with statically linked libraries, and on multiple architectures (in particular we consider AMD64 and ARM as target platforms for our study).
\end{itemize}

The core of \arch is based on recent advancements in the area of natural language processing. 
Specifically, we designed \arch on a Self-Attentive Neural Network recently proposed in \cite{bengioAttentive}.  
The idea is to directly consider the sequence of instructions in the binary function and to model them as natural language.
An GRU Recurrent Neural Network (GRU RNN) is then used to capture the sequential interaction of the instructions. In addition, an attention mechanism allows the final function embedding to consider all the GRU hidden states, and to automatically focus (i.e., give more weight) to the portion of the binary code that helps the most in accomplishing the training objective - i.e., recognise if two binary functions are similar.

We also investigate the possibility of semantically classifying (i.e. identifying the general sematic behavior of) binary functions by clustering similar embeddings. At
the best of our knowledge we are the first to investigate the
feasibility of this task through machine learning tools, and to perform a quantitative analysis
on this subject. The results are encouraging showing a 95\%
of classification accuracy for 4 different broad classes of
algorithms  (namely \emph{Encryption}, \emph{Sorting}, \emph{Mathematical} and \emph{String Manipulation} functions). Finally, we also applied our semantic classifier to known malwares, and we were able to accurately recognize with it functions implementing encryption algorithms.

 \subsection{Contribution} 
The main contributions of our work are:
\begin{itemize}
	\item we describe \arch, a general architecture for calculating binary function embeddings starting from disassembled binaries;
	\item we extensively evaluate \arch showing that it provides better performance than previous state-of-the art systems with similar requirements. Specifically, we compare it with the recent Gemini \cite{CCS}, showing a performance improvement on several metrics that ranges from $6\%$ to $29\%$ depending on the task at hand;
	\item we apply \arch to the problem of identifying vulnerable functions in binary code, a common application task for binary similarity solutions; also in this task \arch provides better performance than state-of-the-art solutions. 
	\item we show that embeddings produced by \arch can be used to automatically classify binary functions in semantic classes. On a dataset of $15$K functions, we can recognize whether a function implements an encryption algorithm, a sorting algorithm, generic math operations, or a string manipulation, with an accuracy of $95\%$.
	\item we apply \arch to the analysis of known malwares, to identify encryption functions. Interestingly, we achieve good performances: among 10 functions flagged by SAFE as \emph{Encryption}, only one was a false positive.
\end{itemize}
The remainder of this paper is organized as follows. Section \ref{sec:related_work} discusses related work, followed by Section \ref{sec:problem_definition} where we define the problem and report an overview of the solution we tested. In Section \ref{sec:theodetails} we describe in details SAFE, and in Section \ref{sec:imptrain} we provide implementation details and information on the training. In Section \ref{sec:evaluation} we describe the experiments we performed and report their results. Finally, in Section \ref{sec;speed} we discuss the speed of SAFE.  

%% file: sec2_relatedwork.tex
\section{Related Work}
\label{sec:related_work}

We can broadly divide the binary similarity literature in works that propose embedding-based solutions, and works that do not.

\subsection{Works not based on embeddings}


\paragraph{Single Platform solutions}\textemdash~Regarding the literature of binary-similarity for a single platform, a family of works is based on matching algorithms for function CFGs. In Bindiff \cite{bindiff} matching among vertices is based on the syntax of code, and it is known to perform poorly across different compiler (see \cite{David}). 
Pewny et al.~\cite{pewny} proposed a solution where each vertex of a CFG is represented with an expression tree; similarity among vertices is computed by using the edit distance between the corresponding expression trees.  

Other works use different solutions that do not rely on on graph matching. David and Yahav~\cite{David2} proposed to represent a function as several independent execution traces, called \emph{tracelets}; similar tracelets are then matched by using a custom edit-distance.  A related concept is used by David et al. in \cite{David} where functions are divided in pieces of independent code, called \emph{strands}. The matching between functions is based on how many statistically significant strands are similar. Intuitively, a strand is significant if it is not statistically common. Strand similarity is computed using an SMT-solver to assess semantic similarity. Note that all previous solutions are designed around matching procedures that work \emph{pair-to-pair}, and they cannot be adapted to pre-compute a constant size signature of a binary function on which similarity can be assessed. 

Egele et al. in \cite{blanket} proposed a solution where each function is executed multiple times in a random environment. During the executions some features are collected and then used to match similar functions. This solution can be used to compute a signature for each function. However, it needs to execute a function multiple times, that is both time consuming and difficult to perform in the cross-platform scenario. Furthermore, it is not clear if the features identified in  \cite{blanket} are useful for cross-platform comparison.   
Finally, Khoo et al.~\cite{khoo} proposed a matching approach based on \emph{n-grams} computed on instruction mnemonics and \emph{graphlets}. Even if this strategy does produce a signature, it cannot be immediately extended to cross-platform similarity. 
 
\begin{table}[t!]
\center
\caption{Notation.}	
\hspace{-1.2cm}
\begin{tabular}{  r | l  }	
  $s$ & source code \\
  $c$ & compiler \\
  $f^{s}$ & binary function compiled from source code $s$. \\
  $\vec{f}$ & embedding vector of function $f$ \\
  $I_{f_i}$ &  list of instructions in function $f_i$ \\
    $m$ & number of instructions in a function  \\
 $\iota$  & instruction in $I_{f_i}$ \\ 
  $\vec{\iota}$  & embedding vector of $\iota$ \\ 
    $\vec{I_{f_i}}$ &  list of instruction embeddings in function $f_i$ \\
  \textsf{i2v} & instruction embedding model (instruction2vector) \\
  $a[i]$ & $i$-th component of vector $a$ \\
  $h_i$ & $i$-th hidden state of the Bi-directional Recurrent Neural Network (RNN) \\
  $n$ & dimension of vector $\vec{f_i}$  \\
    $u$ & dimension of the state $h_i$ \\
  ${\sf tanh}$ & hyperbolic tangent function \\
    ${\sf softmax}$ & softmax function \\
  $\text{ReLU}$ & rectified linear unit function \\
  $W_{s1}$ & $d_a \times u$ weights matrix of the attention mechanism  \\
  $W_{s2}$ & $r \times d_a$ weights matrix of the attention mechanism \\
    $W_{out1}$ & $e \times (m+u)$ weights matrix of the output layer of the Self-Attentive net.  \\
  $W_{out2}$ & $n \times e$ weights matrix of the output layer of the Self-Attentive network \\
$B$ &$r \times u$ embedding matrix   \\
$d_a$ & attention depth - Parameters of the function embedding network\\
$r$ & number of attention hops - Parameters of the function embedding network\\
  $y_i$ & ground truth label associated with the $i$-th input pair of functions \\
  $\Phi$ & network hyper parameters \\
  $J$ & network objective function \\
    $k$ & number of results of a function search query \\
  
\end{tabular}
\label{table:nota}
\end{table}

\paragraph{Cross-Platform solutions}\textemdash~Pewny et al.~\cite{crossarch} proposed a graph-based methodology, i.e. a matching algorithm on the CFGs of functions. The idea is to transform the binary code in an intermediate representation; on such representation the semantic of each CFG vertex is computed by using a sampling of the code executions using random inputs. Feng et al.~\cite{asiaccs} proposed a solution where each function is expressed as a set of conditional formulas; then it uses integer programming to compute the maximum matching between formulas.
Note that both, \cite{crossarch} and  \cite{asiaccs} allow \emph{pair-to-pair} check only. 

David et al.~\cite{David3} propose to transform binary code to an intermediate representation. Then, functions were partitioned in slices of independent code, called \emph{strands}. An involved process guarantees that strands with the same semantics will have similar representations. Functions are deemed to be similar if they have matching of significant strands. Note that this solution does generate a signature as a collection of hashed strands. However, it has two drawbacks: the first is that the signature is not constant-size but it depends on the number of strands contained in the function. The second drawback is that is not immediate to transform such signatures into embeddings that can be directly fed to other machine learning algorithms.

\subsection{Based on embeddings}
The most related to our works are the ones that propose embeddings for binary similarity. Specifically, the works that target cross-platform scenarios.

\paragraph{Single-Platform solutions}\textemdash~Recently, \cite{asm2vec} proposed a function embedding solution called {\em Asm2Vec}. This solution is based on the PV-DM model \cite{LeM14} for natural language processing. Operatively, Asm2Vec computes the CFG of a function, and then it performs a series of random walks on top of it. 
Asm2Vec outperforms several state-of-the-art solutions in the field of binary similarity. Despite being a really promising solution, Asm2vec does not fullfill all the design goals of our system: firstly it requires libc call symbols to be present in the binary code as tokens to produce the embedding of a function; secondly it is only suitable for single-platform embeddings.

\paragraph{Cross-Platform solutions}\textemdash~Feng et al.~\cite{genius} introduced a solution that uses a clustering algorithm over a set of functions to obtain centroids for each cluster. Then, they used these centroids and a configurable feature encoding mechanism to associate a numerical vector representation with each function. Xu et al.~\cite{CCS} proposed an architecture called {\em Gemini}, where function embeddings are computed using a deep neural network. Interestingly, \cite{CCS} shows that Gemini outperforms \cite{genius} both in terms of accuracy and performance (measured as time required to train the model).
In Gemini the CFG of a function is first transformed into an \emph{annotated CFG}, a graph containing manually selected features, and then embedded into a vector using the graph embedding model of \cite{dai2016discriminative}.  The manual features used by Gemini do not need call symbols. At the best of our knowledge Gemini is the state-of-the-art solution for cross-platform embeddings based on deep neural networks that works on cross-platform code without call symbols. In an unpublished technical report \cite{arxiv} we proposed a variation of Gemini where manual features are replaced with an unsupervised feature learning mechanism. This single change led to a $2\%$ performance improvement over the baseline represented by Gemini.

Finally, in \cite{neuralTranslation} the author propose the use of a recurrent neural network based on LSTM (Long short-term memory) to solve a subtask of binary similarity that is the one of finding similar CFG blocks.

\color{black}

%% file: sec3_similaritydefinition.tex

\section{Problem Definition and Solution Overview}
\label{sec:problem_definition}
For clarity of exposition we summarize all the notation used in this paper in Table \ref{table:nota}.
Let us first define the similarity problem. 

We say that two binary functions $f^{s}_1,f^{s}_2$ are similar, $f_1 \sim f_2$, if they are the result of compiling the same original source code $s$ with different compilers. 
Essentially, a compiler $c$ is a deterministic transformation that maps a source code $s$ to a corresponding binary function $f^{s}$. In this paper we consider as a compiler the specific
software, e.g. gcc-5.4.0, together with the parameters that influence the compiling process, e.g.
the optimization flags -O$[0,...,3]$.\\

We indicate with $I_{f_1}:(\iota_1,\iota_2,\iota_3,\ldots,\iota_m)$, the list of assembly instructions composing function $f_1$. 
Our aim is to represent $f_1$ as a vector in $\mathbb{R}^{n}$. This is  achieved with an embedding model that maps $I_{f_1}$ to an {\em embedding vector} $\vec{f_1} \in \mathbb{R}^{n}$, preserving structural similarity relations between binary functions.

{\em Function Semantic.} Loosely speaking, a function $f$ can be seen as an implementation of an algorithm. We can partitions algorithms in {\em classes}, where each class is a group of algorithms solving related problems.  In this paper we focus on four classes $\{${\sf E} (Encryption), {\sf S} (Sorting),  {\sf SM} (String Manipulation), {\sf M} (Mathematical)$\}$. A function belongs to class {\sf E} if it is the implementation of an encryption algorithm (e.g., AES, DES); it belongs to {\sf S} class if it implements a sorting algorithm (e.g., bubblesort, mergesort); 
 it belongs to {\sf SM} class if it implements an algorithm to manipulate a string (e.g., string reverse, string copy);  it belongs to {\sf M} class if it implements math operations (e.g., computing a bessel function); 
 We say that a classifier, recognizes the semantic of a function $f$, with $f$ taken from one of the aforementioned classes, if it is able to guess the class to which $f$ belongs.

\subsection{SAFE Overview.} We use an embedding model structured in two phases; in the first phase the {\it Assembly Instructions Embedding} component, transforms a sequence of assembly instructions $I_f$ in a sequence of vectors, in the second phase a {\it Self-Attentive Neural Network}, transforms a sequence
of vectors in a single embedding vector. See Figure \ref{fig:magg} for a schematic representation of the overall architecture of our embedding network.

\subsubsection*{Assembly Instructions Embedding ({\sf i2v})}
In the first phase of our strategy we map each instruction $\iota \in I_{f}$ to a vector of real numbers $\vec{\iota}$, using the word2vec model \cite{Mikolov}. Word2vec is an extremely popular feature learning technique in natural language processing. We use a large corpus of instructions to train our instruction embedding model (see Section \ref{sec:implementation}), we call our mapping instruction2vec (\textsf{i2v}).
The final outcome of this step is a sequence of vectors $\vec{I_{f}}$.

\subsubsection*{Self-Attentive Network}

For our Self-Attentive Network we use the network recently proposed in \cite{bengioAttentive}. In this Self-Attentive Network, a bi-directional recurrent neural network is fed with the sequence of assembly vectors. Intuitively, for each instruction vector $\vec{\iota_i}$ the RNN computes a summary vector taking into account the instruction itself and its context in $I_{f}$.  The final embedding of $\vec{I_{f}}$ is a weighted sum of all summary vectors. The weights of such summation are computed by a two layers fully-connected neural network.

We selected the Self-Attentive Network for two reasons. First, it shows state-of-the art performance on natural language processing tasks \cite{bengioAttentive}. Secondly, it suffers less of the long-memory problem\footnote{Classic RNNs do not cope well with really long sequences.} of classic RNNs: in the Self-Attentive case the RNN computes only a local summary of each instruction. Our research hypothesis is that it would behave well over the long sequences of instructions composing binary functions; and this hypothesis is indeed confirmed in our experiments (see Section \ref{sec:evaluation}).

%% file: sec4_network.tex
\section{Details of the SAFE, Function Embedding Network}\label{sec:theodetails}

\begin{figure*}[h]
\includegraphics[width=\textwidth]{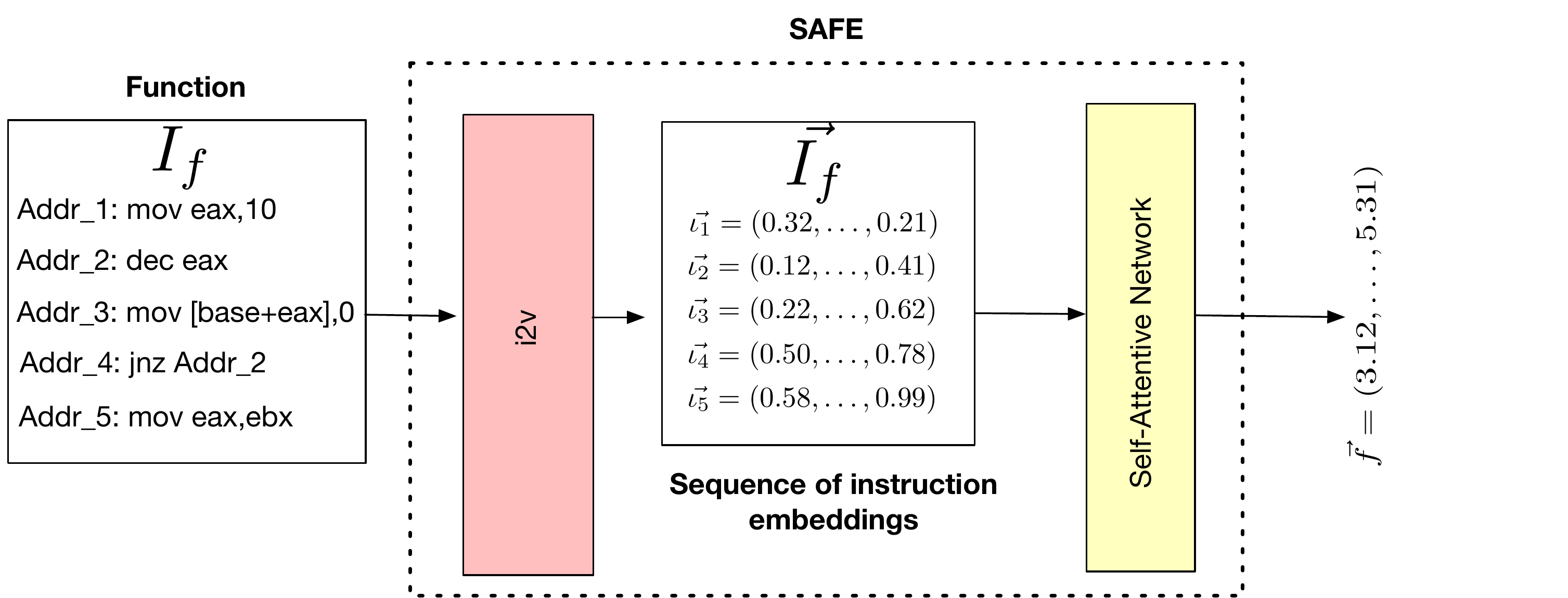}
\caption{Architecture of SAFE. The vertex feature extractor component refers to the Unsupervised Feature Learning case. \label{fig:magg}}
\end{figure*}
We denote the entire embedding network as \archext.

\subsection{Assembly Instructions Embedding ({\sf i2v})}

The first step of our solution consists in associating an embedding vector to each instruction $\iota$ contained in $I_{f}$.
We achieve it by training the embedding model \textsf{i2v} using the skip-gram method \cite{Mikolov}. The idea of skip-gram is to use the current instruction to predict the instructions around it. A similar approach has been used also in \cite{ChuaSSL17}.

We train the  \textsf{i2v} model using assembly instructions as tokens (i.e., a single token includes both the instruction mnemonic and the operands).  We do not use the raw instruction but we filter it as follows.
We examine the operands and replace all base memory addresses with the special symbol \texttt{ MEM} and all immediates whose absolute value is above some threshold (we use $5000$ in our experiments, see Section \ref{sec:implementation}) with the special symbol \texttt{IMM}. 
We do this filtering because we believe that using raw operands is of small benefit; for instance, the displacement given by a jump is useless  (e.g., instructions do not carry with them their memory address), and, on the contrary, it may decrease the quality of the embedding by artificially inflating the number of different instructions.
As example the instruction \texttt{mov EAX,}$6000$ becomes \texttt{mov EAX,IMM}, \texttt{mov EAX,}$[0$\texttt{x}$3435423]$  becomes \texttt{mov EAX,MEM},
while the instruction \texttt{mov EAX,}$[$\texttt{EBP}$-8]$ is not modified. Intuitively, the last instruction is accessing a stack variable different from
\texttt{mov EAX,}[\texttt{EBP}$-4$], and this information remains intact with our filtering.

\subsection{Self-Attentive Network}

\begin{figure*}[h]
\includegraphics[width=0.9\textwidth]{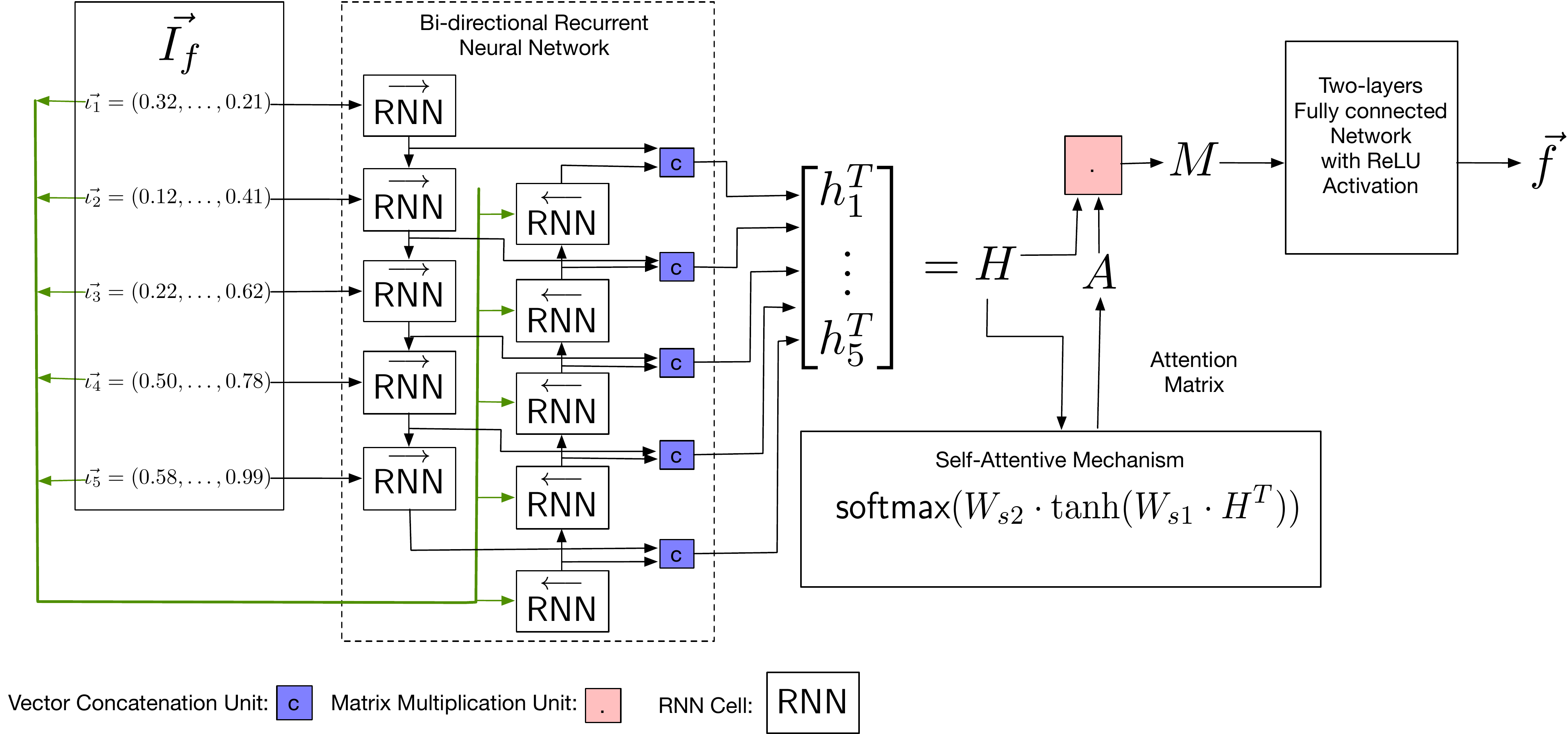}
\caption{Self-Attentive Network: detailed architecture. \label{fig:ssnn}}
\end{figure*}

We based our Self-Attentive Network on the one proposed by \cite{bengioAttentive}. The overall structure is detailed in Figure~\ref{fig:ssnn}. 
We compute embedding $\vec{f}$ of a function $f$ by using the sequence of instruction vectors $\vec{I_{f}}:( \vec{\iota_1},\ldots,\vec{\iota_m})$. These vectors are fed into a bi-directional neural network, 
obtaining for each vector $\vec{\iota_i} \in \vec{I_{f}}$ a {\em summary} vector of size $u$:
$$h_i = \overset{\longrightarrow}{{\sf RNN}} (\overset{\rightarrow}{ h_{i-1}},\iota_i) \oplus    \overset{\longleftarrow}{{\sf RNN}}  (\overset{\leftarrow}{h_{i+1}},\iota_i) $$

\noindent where $\oplus$ is the concatenation  operand,  $\overset{\longrightarrow}{{\sf RNN}}$ (resp., $\overset{\longleftarrow}{{\sf RNN}}$) is the forward (resp., backward) RNN cell, and $\overset{\rightarrow}{h_{i-1}},\overset{\leftarrow}{h_{i+1}}$  are the forward and backward states
of the RNN (we set $\overset{\rightarrow}{h_{-1}}=\overset{\leftarrow}{h_{n+1}}=0$). The state of each RNN cell has size $\frac{u}{2}$.

From these summary vectors we obtain a $m \times u$ matrix $H$.  Matrix $H$ has as rows the summary vectors. 
An attention matrix $A$ of size $r \times m$ is computed using a two layers neural network:
$$A={\sf softmax}(W_{s2} \cdot \tanh(W_{s1}\cdot H^{T}))$$

\noindent where $W_{s1}$ is a weight matrix of size $d_{a} \times u$ and the parameter $d_a$ is the {\em attention depth}  of our model. 
The matrix $W_{s2}$ is a weight matrix of size $r \times d_a$ and the parameter $r$ is the number of {\em attention hops} of our model. 

Intuitively, when $r=1$, $A$ collapses into a single attention vector, where each each value is the weight a specific summary vector. When $r>1$, $A$ becomes
a matrix and each row is an independent attention hop. Loosely speaking, each hops weights the attention of a different aspect of
the binary function. 

The embedding matrix of our sequence is:
$$B=(b_1,b_2,\ldots, b_u)=AH$$
and it has fixed size $r \times u$.
In order to transform the embedding matrix into a vector $\vec{f}$ of size $n$, we flatten the matrix $M$ and we feed the flattening into a two-layers fully connected neural
network with ReLU activation function:
$$\vec{f}=W_{out2} \cdot {\sf ReLU}(W_{out1}\cdot (b_1 \oplus b_2 \ldots \oplus b_u))$$

\noindent where $W_{out1}$ is a weight matrix of size $e \times (r+u)$, and $W_{out2}$
a weight matrix of size $n \times e$.

\bigskip

\begin{figure}[ht]
\center
\includegraphics[width=0.45\textwidth]{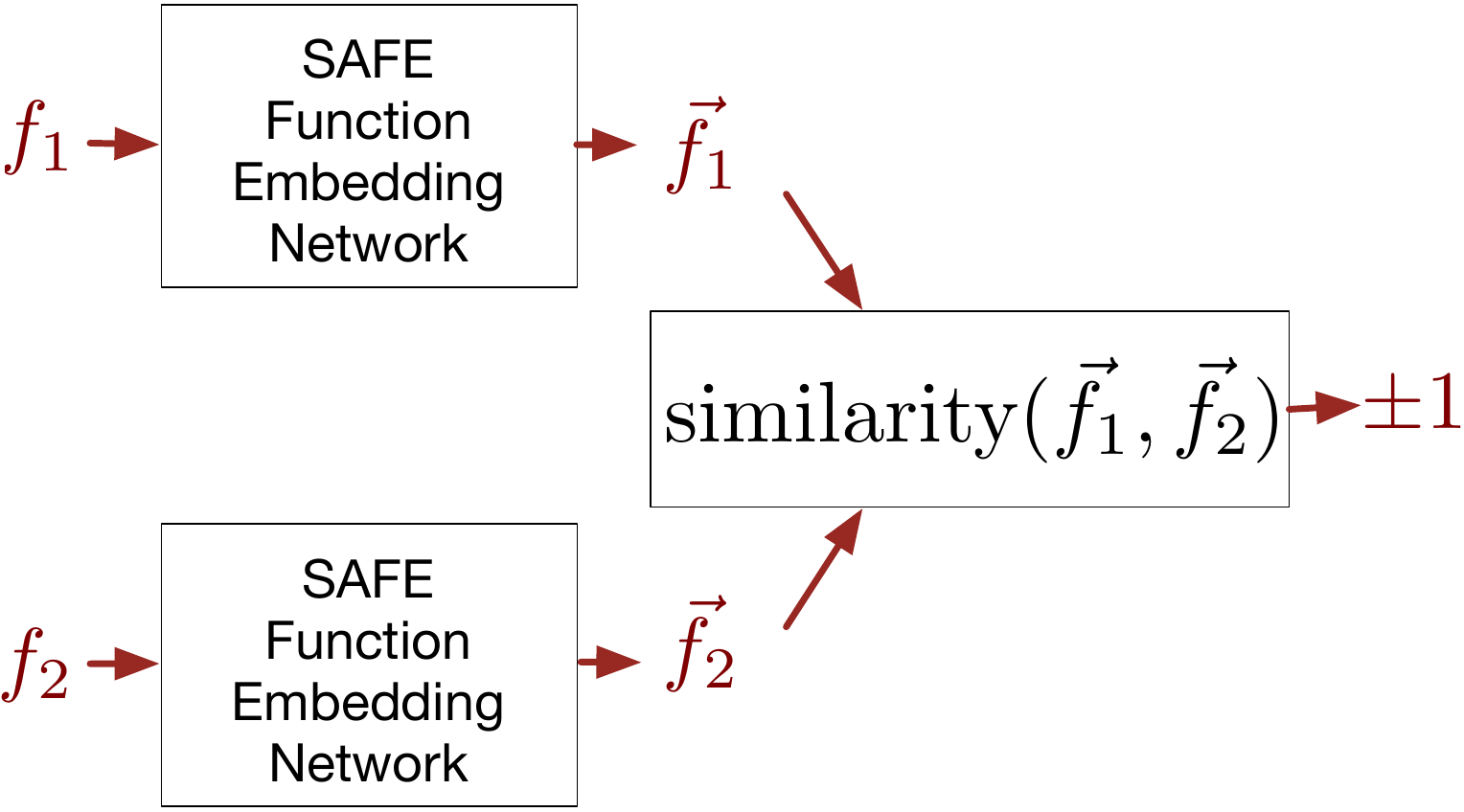}
\caption{Siamese network. \label{fig:siamese}}
\end{figure}

\noindent{{\it Learning Parameters Using Siamese Architecture:}} we learn the network parameters \\$\Phi = \{ W_{s1}, W_{s2},\overset{\longrightarrow}{{\sf RNN}},  \overset{\longleftarrow}{{\sf RNN}} , W_{out1},W_{out2} \}$ using a pairwise approach, a technique also called \emph{siamese network} in the literature \cite{bromley1994signature}.
The main idea is to join two identical function embedding networks with a similarity score (with identical we mean that the networks share the same parameters).
The final output of the siamese architecture is the similarity score between the two input graphs (see Figure \ref{fig:siamese}).

In more details, from a pair of input functions $< f_1, f_2 >$ two vectors $< \vec{f_1}, \vec{f_2} >$ are obtained by using the same function embedding network.
These vectors are compared using cosine similarity as distance metric, with the following formula:
\begin{equation}
	\text{similarity}(\vec{f_1}, \vec{f_2}) = \frac{ \displaystyle\sum_{i=1}^{n} \Big( \vec{f_1}[i] \cdot \vec{f_2}[i]  \Big) }{  \sqrt{  \displaystyle\sum_{i=1}^{n}  \vec{f_1}[i]  } \cdot \sqrt{  \displaystyle\sum_{i=1}^{n}  \vec{f_2}[i]  }  }
\end{equation}
where $\vec{f}[i]$ indicates the $i$-th component of the vector $\vec{f}$.

To train the network we require in input a set of $K$ functions pairs, $< \vec{f_1}, \vec{f_2} >$, with ground truth labels $y_i \in \{+1,-1\}$, where $y_i = +1$ indicates that the two input functions are similar and $y_i = -1$ otherwise. Then using the siamese network output, we define the following objective function:
$$
	J = \displaystyle\sum_{i=1}^{K} \Big(  \text{similarity}(\vec{f_1}, \vec{f_2}) - y_{i}  \Big)^2 + \lVert (A\cdot A^{T}-I) \rVert_{F}
$$

The objective function $J$ is minimized by using, for instance, stochastic gradient descent. The term  $\lVert (A\cdot A^{T}-I) \rVert_{F}$ is introduced to penalize the choice of the same weights for each attention hops in matrix $A$ (see \cite{bengioAttentive}).


%% file: sec5_evaluation.tex

\section{Implementation Details and Training}\label{sec:imptrain}
In this section we first discuss the implementation details of our system, and then we explain how we trained our models. 
\subsection{Implementation Details and i2v setup} \label{sec:implementation}
We developed a prototype implementation of SAFE using Python and the Tensorflow \cite{abadi2016tensorflow} framework. 
For static analysis of binaries we used the ANGR framework \cite{angr},  radare2 \cite{radare2}\footnote{We build our system to be compatible with two opensource leader disassemblers.} and IDA Pro.  
To train the network we used a batch size of 250, learning rate $0.001$, Adam optimizer.

In our SAFE prototype we used the following parameters: the RNN cell is the GRU cell \cite{Gru}; the $u$ value is $100$, $r=10$, $d_a=250$, $e=2000$, $n=100$.

We decided to truncate the number of instructions inside each function to the maximum value of $m=150$, this represents a good trade-off between training time and accuracy, the great majority of functions in our datasets is below this threshold (more than 90\% of the functions). 

 \subsubsection{i2v model}
We trained two \textsf{i2v} models using the two training corpora described below. One model is for the instruction set of ARM and one for AMD64. With this choice we tried to capture the different sintaxes and semantics of these two assembly languages.
The model that we use for \textsf{i2v} (for both versions AMD64 and ARM) is the skip-gram implementation of word2vec provided in \cite{w2vimp}.
We used as parameters: embedding size 100, window size $8$ and word frequency $8$. 
\paragraph{Datasets for training the i2v models}
\label{dataset:i2v}
We collected the assembly code of a large number of functions, and we used it to  build two training corpora, one for the \textsf{i2v} AMD64 model and one for the \textsf{i2v} ARM model. 
We built both corpora by dissasembling several UNIX executables and libraries using IDA PRO.  The libraries and the executables have been randomly sampled from repositories of Debian packages. 

We avoided multiple inclusion of common functions and libraries by using a duplicate detection mechanism; we tested the uniqueness of a function computing an hash of all function instructions, where instructions are filtered by replacing the operands containing immediate and memory locations with a special symbol.

From 2.52 GBs of AMD64 binaries we obtained the assembly code of 547K unique functions. 
From 3.05 GBs of ARM binaries we obtained the assembly code of 752K unique functions.
Overall the AMD64 corpus contains $86$ millions assembly code lines while the ARM corpus contains $104$ millions assembly code lines.

\subsection{ Training Single and Cross Platform Models}\label{sec:training}
We trained SAFE models using the same methodology of Gemini, see \cite{CCS}. 
We trained both a single and a cross platform models that were then evaluated in several tasks (see Section \ref{sec:evaluation} for the results).

\subsubsection{Datasets}\label{sec:ourdatasets}

\begin{itemize}
\item {\bigDataset}.
This is dataset has been obtained by compiling the following libraries for AMD64: binutils-2.30, ccv0.7, coreutils-8.29, curl-7.61.0, gsl-2.5, libhttpd-2.0, openmpi-3.1.1, openssl-1.1.1-pre8, valgrind-3.13.0. The compilation has been done using 3 different compilers, clang-3.9, gcc-5.4, gcc-3.4\footnote{Note that gcc-3.4 has been released  more than 10 years before gcc-5.4.}  and 4 optimization levels (i.e., -O$[0$-$3]$). The compiled object files have been disassembled with ANGR, obtaining a total of 452598 functions. 

\item {\normalDataset}.
To align our experimental evaluation with state-of-the-art studies we built the \normalDataset in the same way as the one used in \cite{CCS}. In particular, the \normalDataset consists of a set of 95535 functions generated from all the binaries included in two versions of Openssl (v1\_0\_1f - v1\_0\_1u) that have been compiled for AMD64 and ARM using gcc-5.4 with 4 optimizations levels (i.e., -O$[0$-$3]$). The resulting object files have been disassembled using ANGR; we discarded all the functions that ANGR was not able to disassemble.

\end{itemize}
\subsubsection{Training}
We generate our training and test pairs as reported in \cite{CCS}.
The pairs can be of two kinds: similar pairs, obtained pairing together two binary functions originated by the same source code, and dissimilar pairs, obtained pairing randomly functions that do not derive from the same source code. 
 
Specifically, for each function in our datasets we create two pairs, a similar pair, associated with training label $+1$ and a dissimilar pair, training label $-1$; obtaining a total number of pairs that is twice the total number of functions.

The functions in \bigDataset are partitioned in three sets: train, validation,  and test (75\%-15\%-15\%).  

The functions in \normalDataset are partitioned in two sets: train and test (80\%-20\%), in this case we do not need the validation set because in Task 1 Section \ref{task1} we will perform a cross-validation.

The test and validation pairs will be used to assess  performances in Task 1, see Section \ref{task1}.

As in \cite{CCS}, pairs are partitioned preventing that two similar functions are in different partitions (this is done to avoid that the network sees during training functions similar to the ones on which it will be validated or tested). 

We train our models for $50$ epochs (an epoch represents a complete pass over the whole training set). In each epoch we regenerate the training pairs, that is we create new similar and dissimilar pairs using the functions contained in the training split. We pre-compute the pairs used in each epoch, in such a way that each method is tested on the same data. 
Note that, we do not regenerate the validation and test pairs. 

\section{Evaluation}\label{sec:evaluation}
We perform an extensive evaluation of SAFE investigating its performances on several tasks:
\begin{itemize}
\item {\bf Task 1 - Single Platform and Cross Platform Models Tests}:  we test our single platform and cross platform models following the same methodology of \cite{CCS}. We achieve a performance improvement of $6.8\%$ in the single platform case and of $4.4\%$ in the cross platform case. We remark that in these tests our models behave almost perfectly (within $1\%$ from what a perfect model may achieve). This task is described in Section \ref{task1}.
\item {\bf Task 2 - Function Search}: in this task we are given a certain binary function and we have to search for similes on a large dataset created using several compilers (including compilers that were not used in the training phase). 
We achieve a precision above $80\%$ for the first 15 results, and a recall of $47\%$ in the first 50 results. 
Section \ref{task2} is devoted to Task 2. 
\item {\bf Task 3 - Vulnerability Search}: in this task we evaluate our system on a use-case scenario in which we search for vulnerable functions.  Our tests on several vulnerabilities show  a recall of $84\%$ in the first 10 results.  Task 4 is the focus of Section \ref{task3}. 

\item {\bf Task 4 - Semantic Classification}: in this task we  classify the semantic of binary functions using the embeddings built with SAFE. We reach an accuracy of $95\%$ on our test dataset. Moreover,  we test our classifier on real world malwares, showing that we can identify encryption functions. Task 4 is explained in Section \ref{task4}. 

\end{itemize}

During our evaluation we compare safe with Gemini \footnote{Gemini has not been distributed publicly. We implemented it using the information contained in \cite{CCS}.  For Gemini the parameters are: function embeddings of dimension $64$, number of rounds $2$, and a number of layers $2$. These parameters are the ones that give the better performance for Gemini, according to 
our experiments and the one in the original Gemini paper. }

\subsection{Task 1 - Single and Cross Platform tests}\label{task1}
In this task we evaluate the performance of SAFE using the same testing methodology of Gemini. 
We use the test split and the validation split computed as discussed in Section \ref{sec:ourdatasets}.

\subsubsection{Test methodology}
We perform two disjoint tests. 
\begin{itemize}
\item On \bigDataset, we compute performance metrics on the validation set for all the epochs. Then, we use the model hyper parameters that led to the best performance on the validation set to compute a final performance score on the test set. 

\item On \normalDataset, we perform a 5-fold cross validation: we partition the dataset in $5$ sets; for all possible set union of 4 partitions we train the classifiers on such union and then we test it on the remaining partition. The reported results are the average of 5 independent runs, one for each possible fold chosen as test set. This approach is more robust than a fixed train/validation/test split since it reduces the variability of the results. 
\end{itemize}

\paragraph{Measures of Performances} 
As in \cite{CCS}, we measure the performance using the  \emph{Receiver Operating Characteristic} (ROC) curve \cite{herlocker2004evaluating}. Following the best practices of the field we measure the {\em area under the ROC curve}, or AUC (Area Under Curve). Loosely speaking, higher the AUC value, better the predictive performance of the algorithm. 

\subsubsection{Experimental Results}

\begin{figure*}[ht!]
\center

\subfloat[Test on \bigDataset. ROC curves for the comparison between SAFE and Gemini on the test set. The dashed line is the ROC for Gemini, the continuous line the ROC for SAFE. The AUC of our solution is $0,990$ the AUC of Gemini is $0,932$\label{fig:roc_curve1} ]{\includegraphics[width=.48\textwidth]{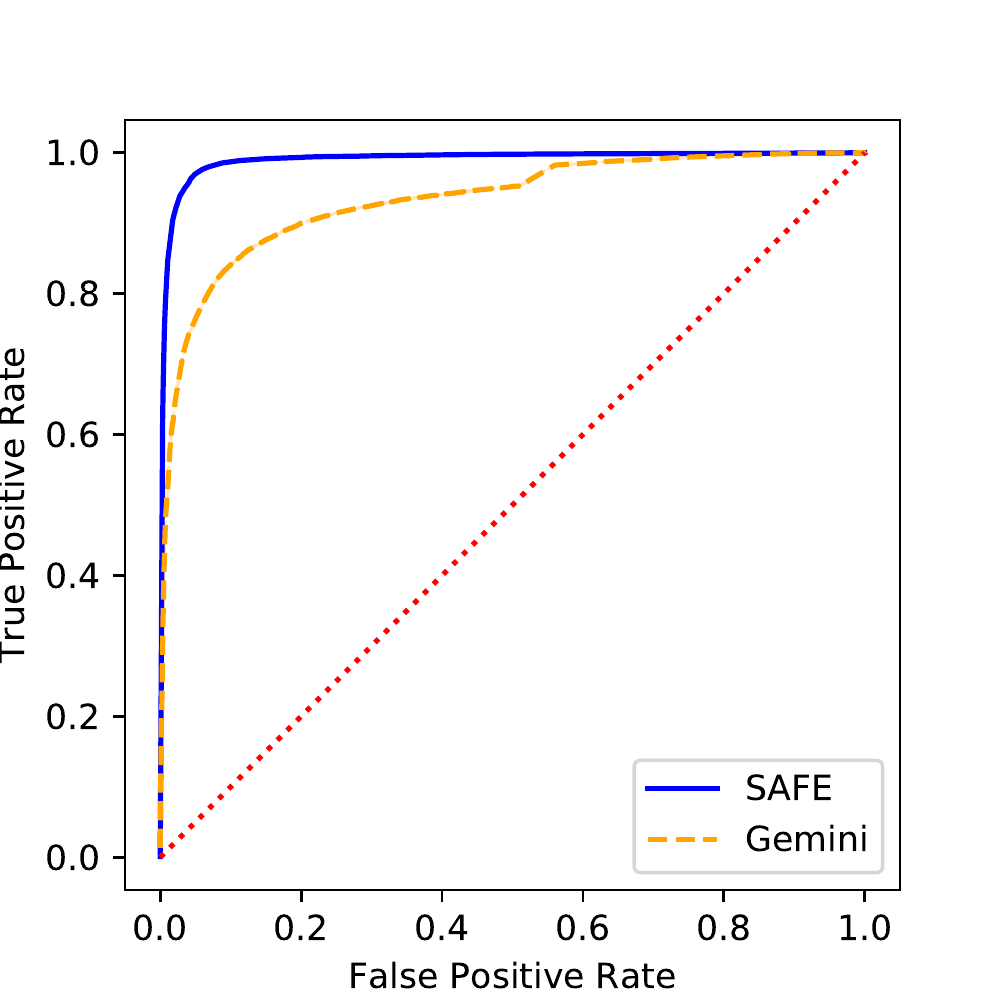}}\hfill
\subfloat[Test on \normalDataset. ROC curves for the comparison between SAFE and Gemini using 5-fold cross validation. The lines represent the ROC curves obtained by averaging the results of the five runs; the dashed line is the average for Gemini, the continuous line the average for our solutions. For both we color the area between the ROC curves with minimum AUC and the maximum AUC. The average AUC of our solution is $0,992$ the average AUC of Gemini is $0,948$\label{fig:roc_curve2} ]{\includegraphics[width=0.48\linewidth]{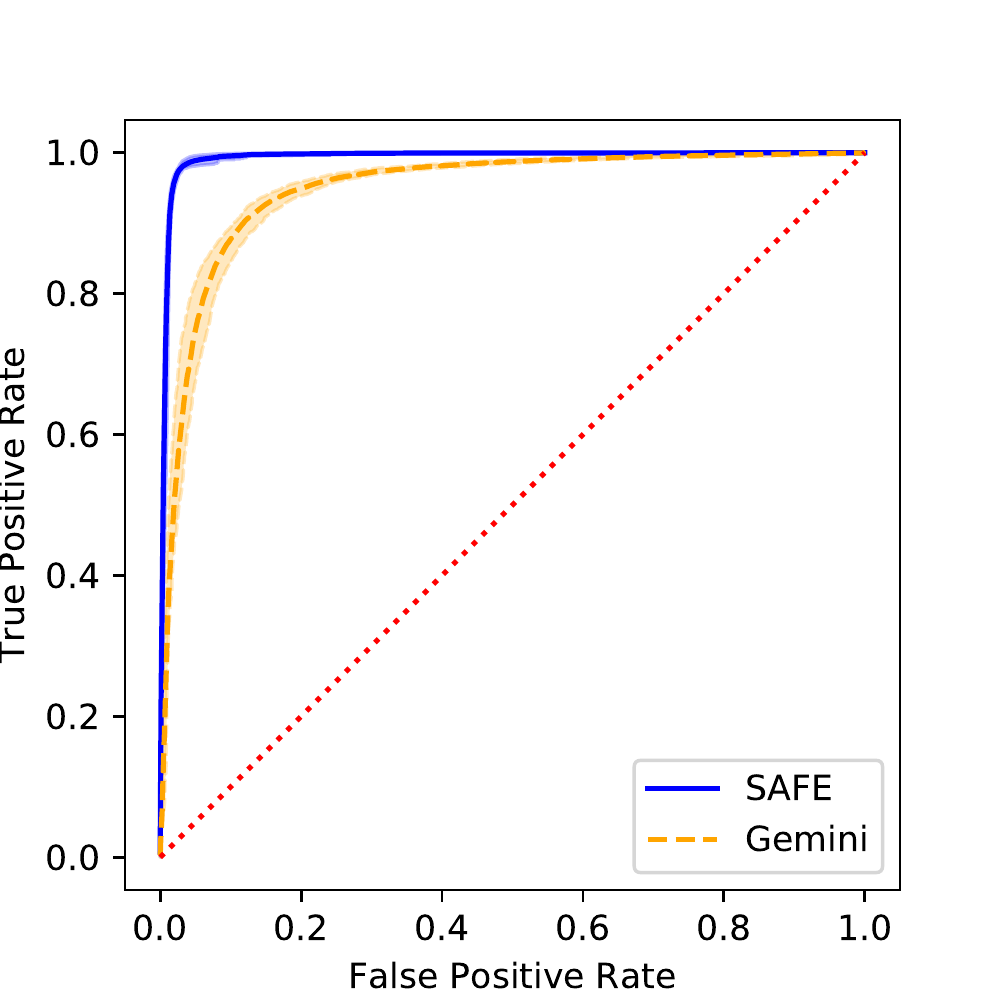}}\hfill
\caption{ROC curves on \bigDataset and \normalDataset. Task 1- Validation and Test of Single Platform and Cross Platform Models}\label{fig:roc_curves}
\end{figure*}

\paragraph{\bigDataset} The results for the single platform case are in Figure \ref{fig:roc_curve1}.   Our AUC is $0.99$, the AUC of Gemini is $0.932$. Even if the improvement is 6.8, it is worth to notice that SAFE provides performance that are close to the perfect case (0.99 AUC).

\paragraph{\normalDataset}

We compare ourselves with Gemini in the cross-platform case. The results are in Figure \ref{fig:roc_curve2} and they  shows the average ROC curves on the five runs of the $5$-fold cross validation. The Gemini results are reported with an orange dashed line while we use a continuous blue line for our results. For both solutions we additional highlighted the area between the ROC curves with minimum AUC maximum AUC in the five runs.
The better prediction performance of SAFE is clearly visible; the average AUC obtained by Gemini is $0,948$ with a standard deviation of $0,006$ over the five runs, while the average AUC of SAFE is $0,992$ with a standard deviation of $0,002$.  
The average improvement  with respect to Gemini is of $4.4\%$.

\subsection{Task 2 - Function Search}\label{task2}

In this task we evaluate the function search capability of the model trained on \bigDataset. We take a target function $f$, we compute its embedding $\vec{ f}$ and we search for similar functions in the \postgres (details of this dataset are given below). 
Given the target $\vec{f}$, a search query returns $R_{\vec{f}}:(r_1,r_2,\dots,r_k)$, that is the ordered list of the $k$ nearest embeddings in \postgres. 

\subsubsection{Dataset}
We built \postgres by compiling postgreSQL 9.6.0 for AMD64 using 12 compilers: gcc-3.4, gcc-4.7, gcc-4.8,  gcc-4.9, gcc-5.4, gcc-6, gcc-7, clang-3.8, clang-3.9, clang-4.0, clang-5.0, clang-6.0. For each compiler we used all 4 optimization levels. We took the object files, i.e. we did not create the executable by linking objects file together, and we disassembled them with radare2, obtaining a total of 581640 functions. For each function the \postgres contains an average number of $33$ similars. We do not reach an average of $48$\footnote{48= 12 compilers $\times$ 4 optimizations level} similars because some functions are lost due to disassembler errors.

\subsubsection{Measures of Performances} 
We compute the usual measures of {\em precision},  fraction of similar functions in $R_{\vec{f}}$ over all functions in $R_{\vec{f}}$, and {\em recall}, fraction of similar functions in $R_{\vec{f}}$ over all similar functions in the dataset. 
Moreover, we also compute the {\em normalised Discounted Cumulative Gain (nDCG)} \cite{dcg}:
$$nDCG(R_{\vec{f}}) = \frac{\sum^{k}_{i=1}\frac{isSimilar(r_i,\vec{f})}{\log(1+i)}}{IdealDCG_{k}} $$
Where $isSimilar$ is $1$ if $r_i$ is a function similar to $\vec{f}$ or $0$ otherwise, and, $IdealDCG_{k}$ is the Discounted Cumulative Gain of the optimal query answering. This measure is between $0$ and $1$, and it takes into account the ordering of the similar functions in $R_{\vec{f}}$, giving better results to responses that put similar functions first. 

As an example let us suppose we have two results for the same query: $(1,1,0,0)$ and $(1,0,0,1)$ (where $1$ means that the corresponding index in the result list is occupied by a similar function and $0$ otherwise). These results have the same precision (i.e., $\frac{1}{2}$), but nDCG scores the first better. 

\subsubsection{Experimental Results}
\begin{figure*}[t!]
\centering
\subfloat[Precision for the  top-$k$ answers with $k \leq 50$. \label{fig:precision} ]{\includegraphics[width=.33\textwidth]{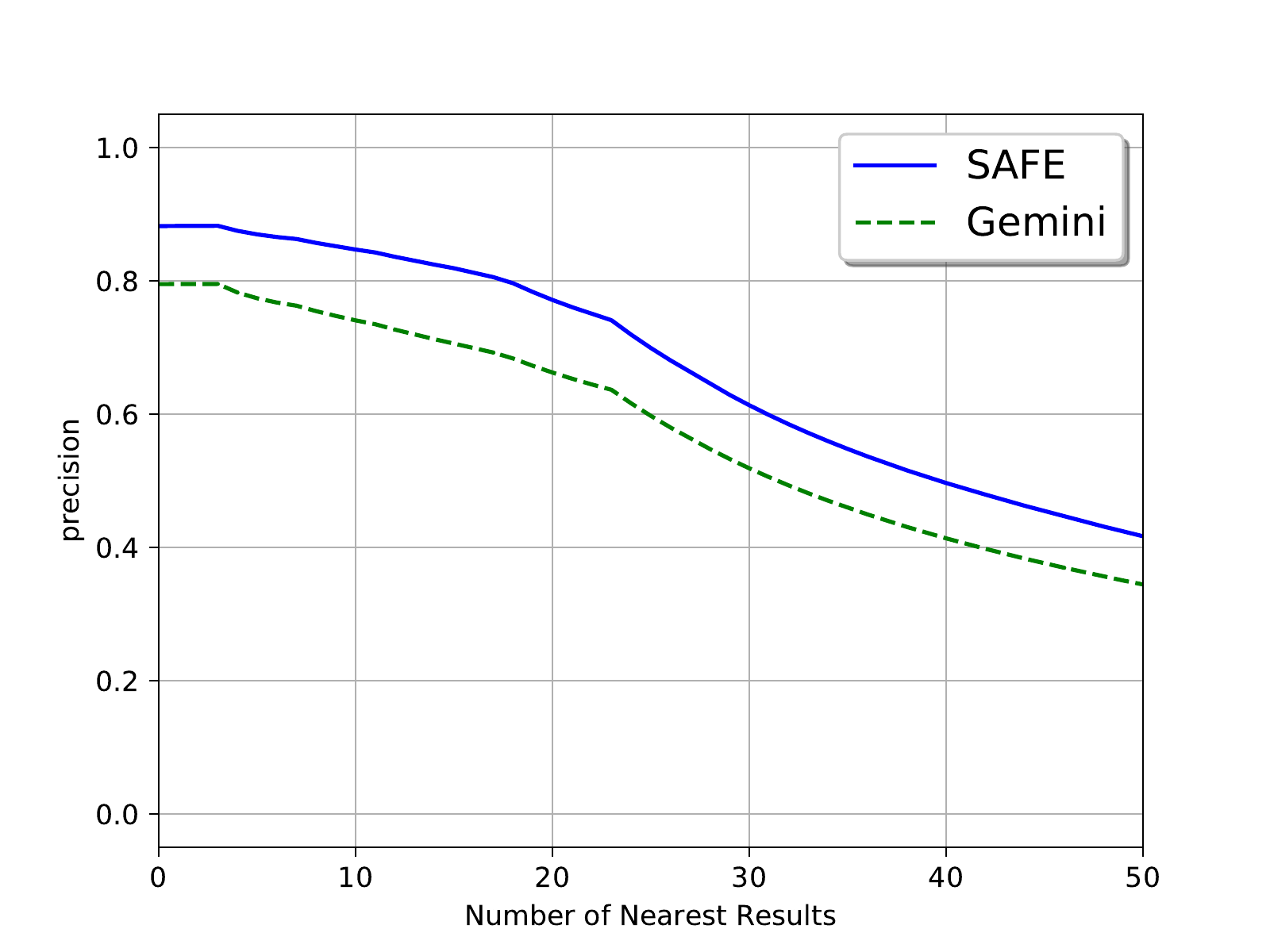}}\hfill
\subfloat[nDCG for the top-$k$ answers with $k \leq 50$. \label{fig:ndcg} ]{\includegraphics[width=.33\textwidth]{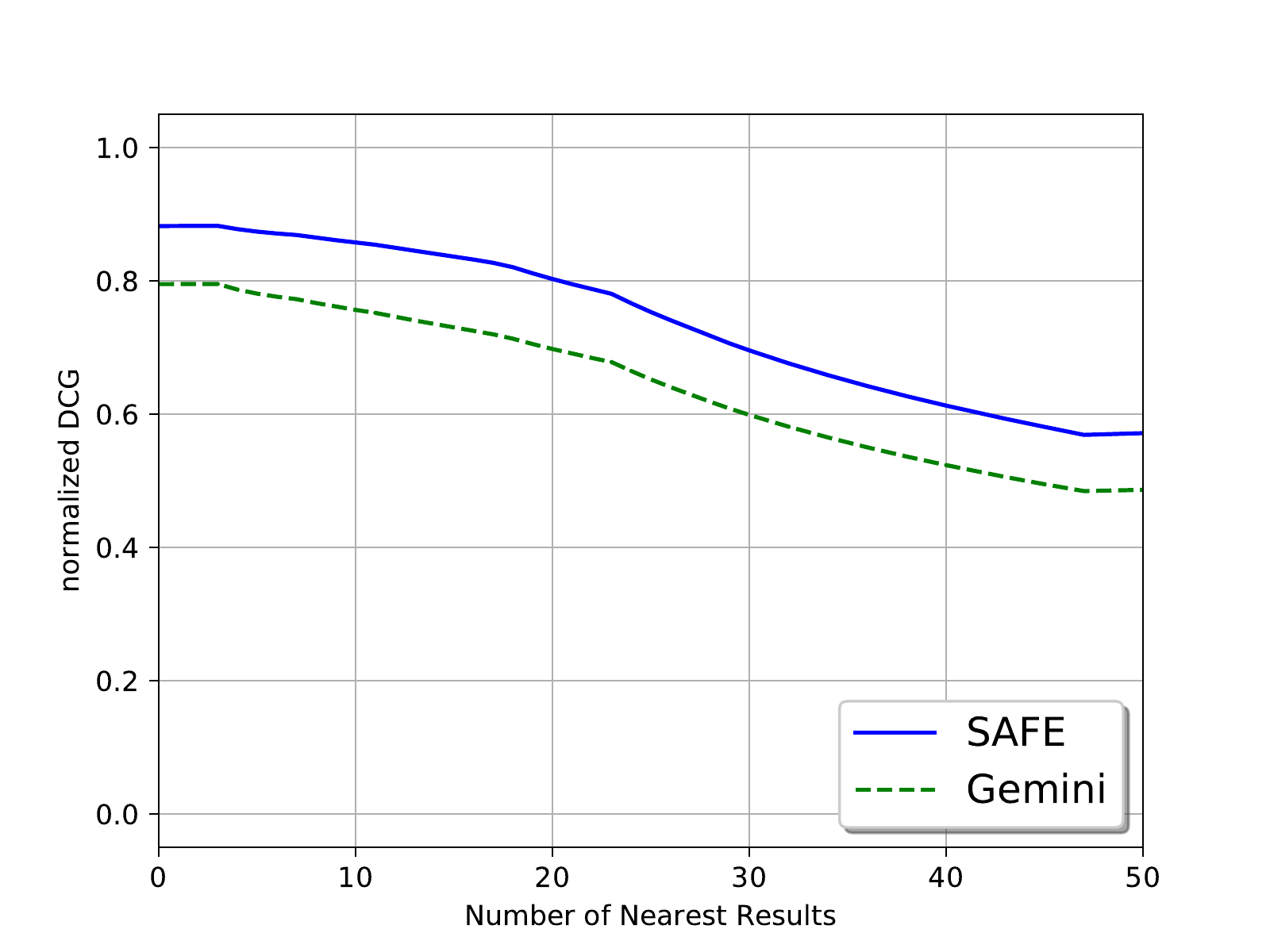}}\hfill
\subfloat[Recall for the  top-$k$ answers with $k \leq 200$. \label{fig:recall} ]{\includegraphics[width=.33\textwidth]{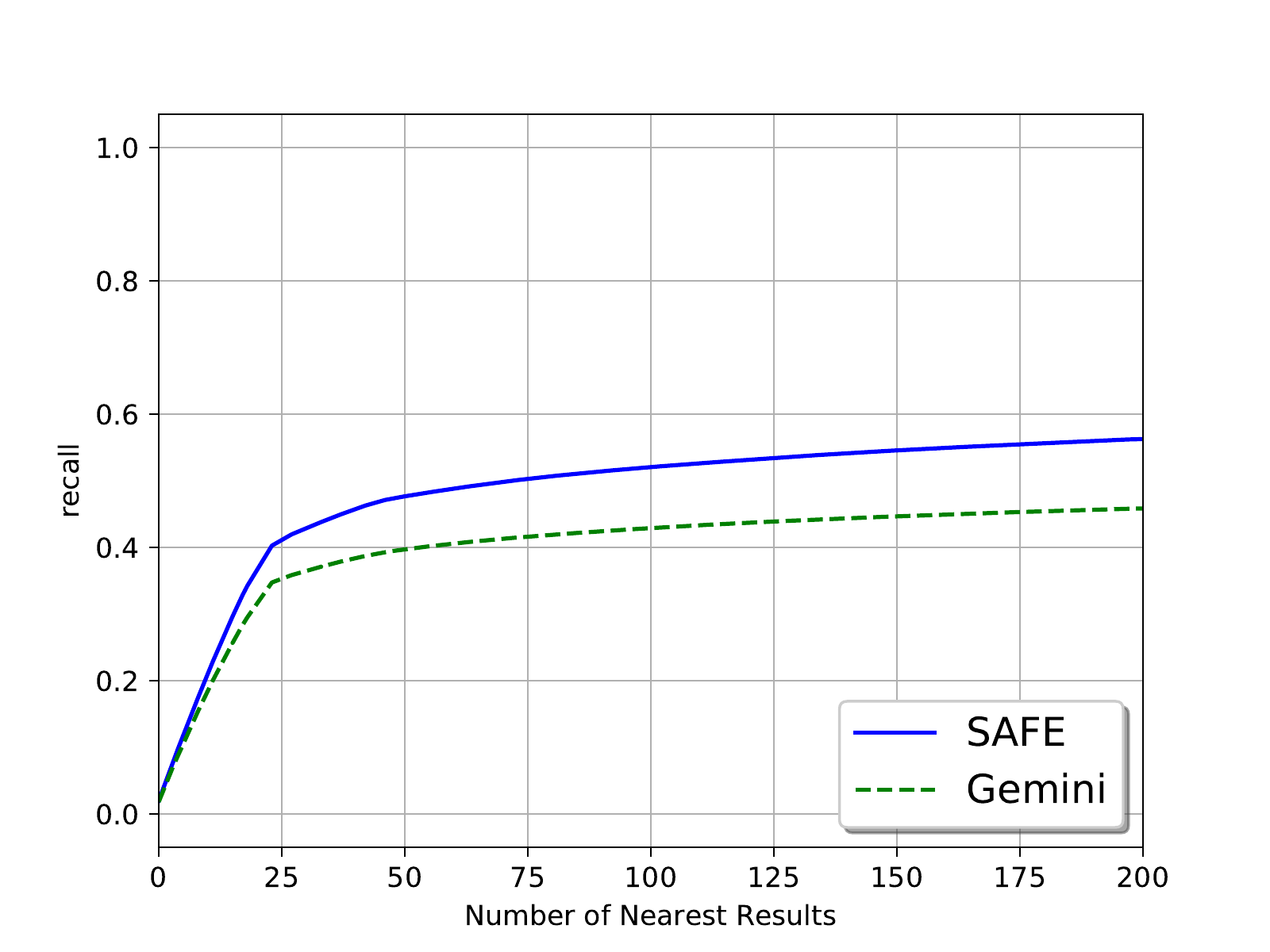}}\hfill
\caption{Results for Task 3 -Function Search, on \postgres (581K functions) average on 160K queries. \label{fig:funcsearch}}
\end{figure*}

Our results on precision, nDCG and recall are reported in Figure \ref{fig:funcsearch}.

 The performances were calculated by averaging the results of 160K  queries. The queries are obtained by sampling, in \postgres, 10K functions 
for each compiler and optimization level in the set $\{$clang-4.0,clang-6.0,gcc-4.8,gcc-7$\} \times \{$O0,O1,O2,O3$\}$.

  Let us recall that, on average, for each query we have 33 similar functions (e.g., functions compiled from the same source code) in the dataset. 
\begin{itemize}
\item Precision:  The results are reported in Figure \ref{fig:precision}. The precision is above $80\%$ for $k \in [0,15]$, and it is above $60\%$ for $k \in [0,30]$. The increase of performance on Gemini is around 10\% on the entire range considered. 
Specifically at $k \in \{10,20,30,40,50\}$ we have values $\{84\%,77\%,61\%,49\%,41\%\}$ for SAFE and $\{74\%,66\%,51\%,41\%,34\%\}$ for Gemini. 

\item nDCG: The tests are reported in Figure \ref{fig:ndcg}. Our solution has a performance above $80\%$ for $k \in [0,18]$. This implies that we have a good order of the results and
the similar functions are among the first results returned. 
The value is always above $50\%$. There is a clear improvement with respects to Gemini, the increase is around $10\%$ on the entire range considered. 
Specifically at $k \in \{10,20,30,40,50,100,200\}$ we have values $\{85\%,80\%,69\%,61\%,57\%,59\%,62\%\}$ for SAFE and $\{75\%,69\%,59\%,52\%,48\%, 50\%, 52\%\}$ for Gemini.

\item Recall:  The tests are reported in Figure \ref{fig:recall}. We have a recall at $k=50$ of 47\% (vs. 39\% Gemini), the recall at $k=200$ is $56\%$ (vs. 45\% Gemini). 
Specifically at $k \in \{10,20,30,40,50,100,200\}$ we have values $\{21\%,36\%,42\%,45\%,47\%,52\%,56\%\}$ for SAFE and $\{18\%,31\%,36\%,38\%,39\%, 42\%, 45\%\}$ for Gemini. 

\end{itemize}

\subsection{Task 3 - Vulnerability Search}\label{task3}
In this task we evaluate our ability to look up for vulnerable functions on a dataset specifically designed for this purpose. The methodology and the performance measures of this test are the same of Task 2. 

\subsubsection{Dataset and methodology} 
The dataset used is the vulnerability dataset of \cite{David}. The dataset contains several vulnerable binaries compiled with 11 compilers in the families of clang, gcc and icc.
The total number of different vulnerabilities is 8\footnote{cve-2014-0160, cve-2014-6271, cve-2015-3456, cve-2014-9295, cve-2014-7169, cve-2011-0444, cve-2014-4877, cve-2015-6862.}. We disassembled the dataset with ANGR,  obtaining  3160 binary functions. 
The average number of vulnerable functions for each of the 8 vulnerabilities is $7,6$; with a minimum of $3$ vulnerable functions and a maximum of $13$\footnote{Some vulnerable functions are lost during the disassembling process}.
We performed a lookup for each of the 8 vulnerabilities, computing  the precision, nDCG, and recall on each result. Finally, we averaged these performances over the
8 queries. 
 
\subsubsection{Experimental Results}
\begin{figure*}[t!]

\centering
\subfloat[Precision for the  top-$k$ answers with $k \leq 50$. \label{fig:precisionVULN} ]{\includegraphics[width=.33\textwidth]{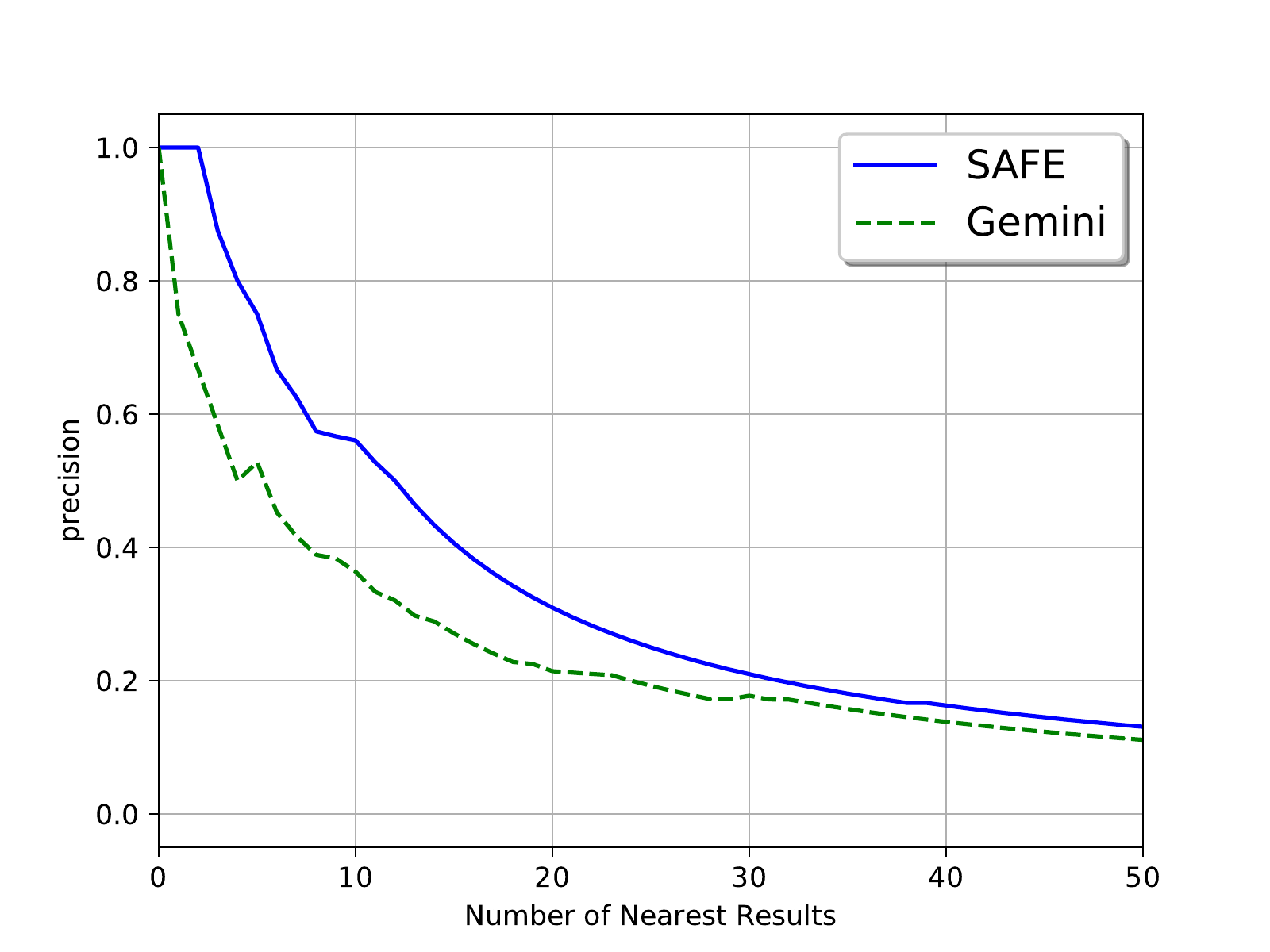}}\hfill
\subfloat[nDCG for the top-$k$ answers with $k \leq 50$. \label{fig:ndcgVULN} ]{\includegraphics[width=.33\textwidth]{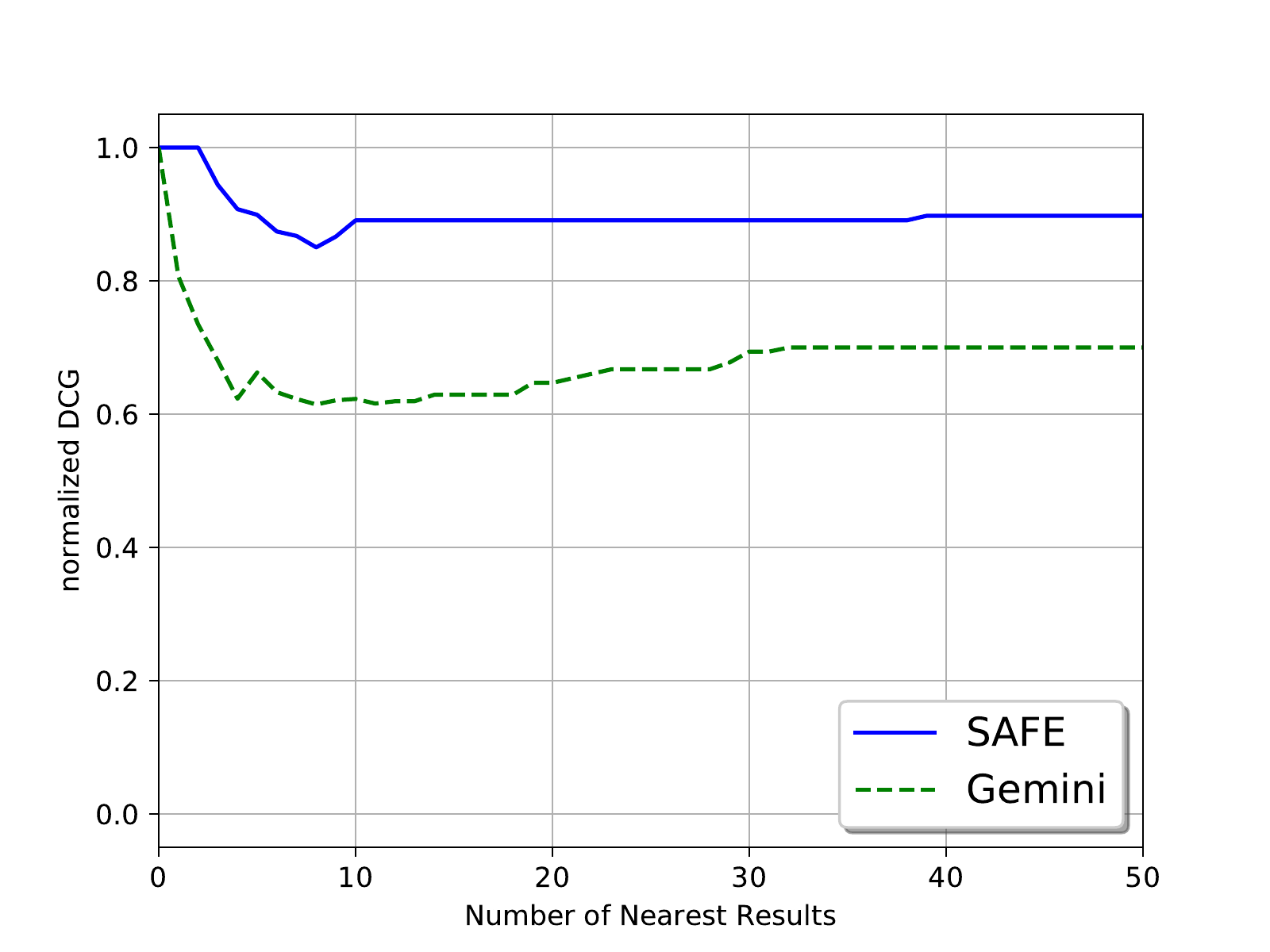}}\hfill
\subfloat[Recall for the  top-$k$ answers with $k \leq 200$. \label{fig:recallVULN} ]{\includegraphics[width=.33\textwidth]{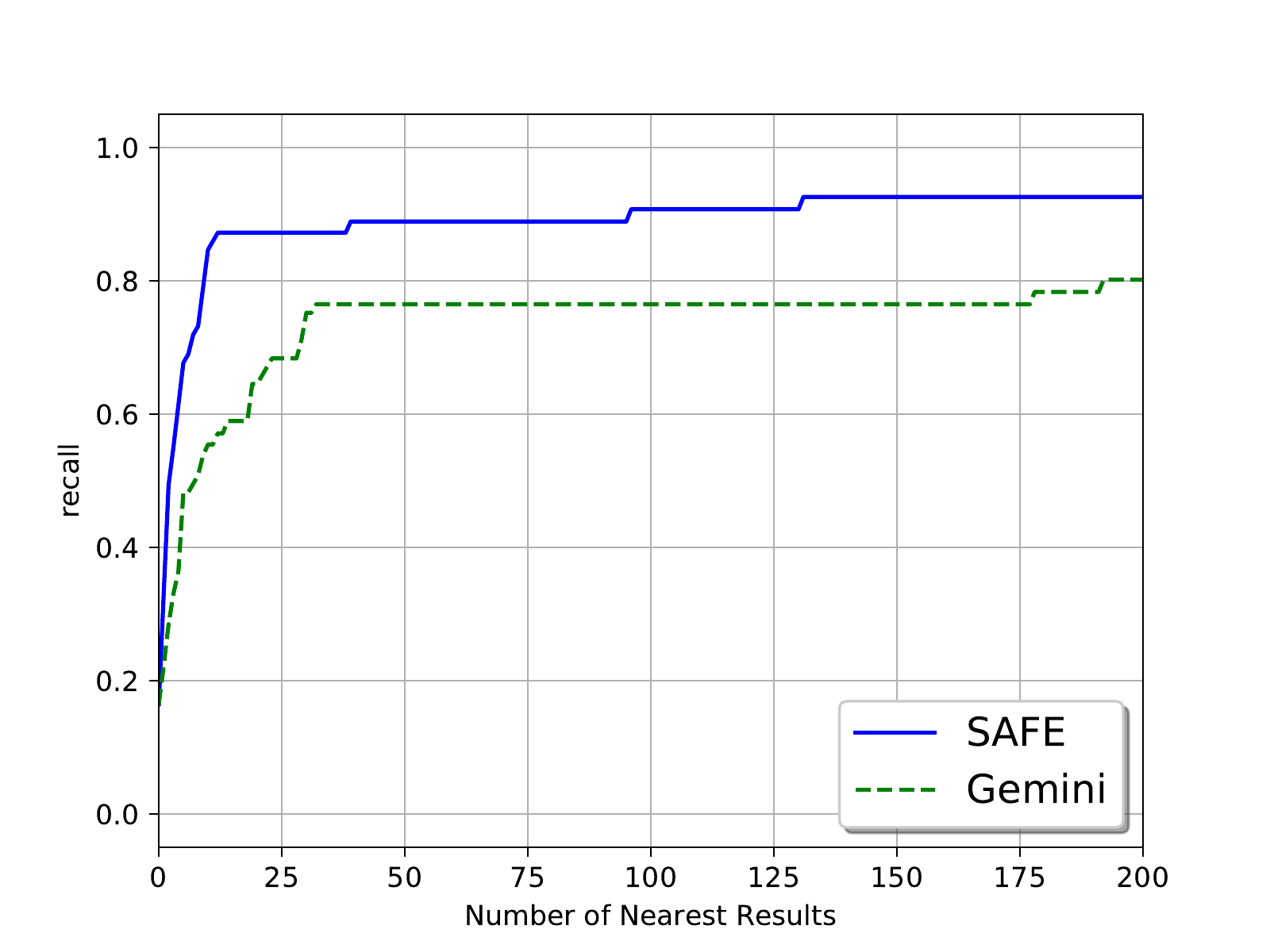}}\hfill
\caption{Results for Task 3 - Vulnerability Search. \label{fig:vulnsearch}}
\end{figure*}

The results of our experiments are reported in Figure \ref{fig:vulnsearch}. We can see that SAFE outperforms Gemini for all values of $k$ in all tests.  Our nDCG is very large, showing that SAFE effectively finds  most of the vulnerable functions in the nearest 
results. 
For $k=10$ we reach a recall of $84\%$, while Gemini reaches a recall of $55\%$. For $k=15$ our recall is $87\%$ (vs. $58\%$ recall of Gemini, with an increment of performance of 29\%), and we reach a maximum of $88\%$ (vs. $76\%$ of Gemini).
One of the reason why the accuracy quickly decreases is that, on average, we have $7,6$ similar functions; this means that even a perfect system at $k=20$ will have an accuracy that is less than $50\%$.
This metric problem is not shared by the nDCG reported in Figure \ref{fig:ndcgVULN}, recall that the nDCG is normalized on the behaviour of the perfect query answering system. 
During our tests we have seen that on the infamous hearthbleed vulnerability we have an ideal behaviour, SAFE found all the $13$ vulnerable functions in the first $13$ results, while Gemini had a recall at 13 around $60\%$. 

\subsection{Task 4 - Semantic Classification}\label{task4}
In Task 4 we evaluate the semantic classification using the embeddings computed with the model trained on \bigDataset. 
We calculate the embeddings for all functions in \simDataset (details on the dataset below). We split our embeddings in train set and test set and we train and test an SVM classifier using a 10-fold cross validation.  We use an SVM classifier with kernel rbf, and parameters $C=10$ and $\gamma=0.01$. 
We compare our embeddings with the ones computed with Gemini.

 \begin{table}
	\centering
	\caption{Number of function for each class in the \simDataset 	\label{table:semanticdataset}
}
	\begin{tabular}{| r | c |}
	\hline
	\textbf{Class} & \textbf{Number of functions} \\
	\hline
   {\sf S} (Sorting) & 4280 \\	
  {\sf E} (Encryption) & 2868 \\
  {\sf SM} (String Manipulation) & 3268 \\
  {\sf M}  (Math) & 4742 \\
	\hline
	\textbf{Total} &  15158 \\
	\hline
	\end{tabular}
\end{table}
\subsubsection{Dataset} The \simDataset has been generated from a source code collection containing 443 functions that have been manually annotated as implementing algorithms in one of the 4 classes: {\sf E} (Encryption), {\sf S} (Sorting),  {\sf SM} (String Manipulation), {\sf M} (Mathematical). \simDataset contains multiple functions that refer to different implementations of the same algorithm.  We compiled the sources for AMD64 using the 12 compilers and 4 optimizations used for \postgres, we took the object files and after disassembling them with ANGR we obtained a total of 15158 binary functions, see details in Table \ref{table:semanticdataset}. It is customary to use auxiliary functions when implementing complex algorithms (e.g. a swap function used by a quicksort algorithm). When we disassemble the \simDataset we take special care to include the auxiliary functions in the assembly code of the caller.   This step is done to be sure that the semantic of the function is not lost due to the scattering of the algorithm semantic among helper functions.  Operatively, we include in the caller all the callees up to depth 2. 

\subsubsection{Measures of Performances} 
 As performance measures we use precision, recall and F-1 score.

\subsubsection{Experimental Results} 


\begin{table}[ht!]
		\center
		\caption{Results of semantic classification using embeddings computed with SAFE model and Gemini. The classifier is an SVM with kernel \textit{rbf}, $C=10$ and $gamma=0.01$. \label{table:Semantic_classification_result}}

	\begin{tabular}{ p{1.4cm} | c | c | c  | c |}
		\textbf{Class} & \textbf{Embedding Model} & \textbf{Precision} & \textbf{Recall} & \textbf{F1-Score} \\
		\hline
		\multirow{2}{*}{  \makecell{{\sf E} \\(Encryption)}     } &  \textbf{SAFE} &    \textbf{0.92}   &    \textbf{0.94}    &    \textbf{0.93}    \\ 		\cline{2-5}
													& Gemini  &    0.82   &    0.85    &    0.83   \\

		\hline
		\multirow{2}{*}{  \makecell{{\sf M}  \\ (Math.)}  }      & \textbf{SAFE} &     \textbf{0.98}  &    \textbf{0.95}   &    \textbf{0.96}		\\
		\cline{2-5}
													  	 & Gemini &     0.96  &    0.90   &    0.93     \\
		\hline
        \multirow{2}{*}    {  \makecell{{\sf S} \\(Sorting)}}			  & \textbf{SAFE} &		\textbf{0.91}  &    \textbf{0.93}   &    \textbf{0.92}   \\     
        \cline{2-5}
        											 & Gemini &		0.87  &     0.92   &    0.89   \\    
        \hline
      	\multirow{2}{*}{\makecell{  {\sf SM} (String\\ Manipulation)} }    & \textbf{SAFE} &		\textbf{0.98}  &  	 \textbf{0.97}   &   \textbf{0.97}   \\ 
      	\cline{2-5}
      												 & Gemini &		0.90  &  	 0.89   &    0.89   \\
      	\hline
      	\hline
      	
      	\multirow{2}{*}{\makecell{Weighted\\ Average}}  & \textbf{SAFE} &		\textbf{0.95}  &  	 \textbf{0.95}   &   \textbf{0.95} \\      	\cline{2-5}
      															  &  Gemini &  0.89  &    0.89  &    0.89 \\

      	\hline
	
	\end{tabular}	

\end{table}

The results of our semantic classification tests are reported in Table \ref{table:Semantic_classification_result}. First and foremost, we have a strong confirmation that is indeed possible to classify the semantic of the algorithms using function embeddings. 
The use of an SVM classifier on the embedding vector space leads to good performance. There is a limited variability of  performances between
  different classes. The classes on which SAFE performs better are {\sf SM} and {\sf M}. We speculate that the moderate simplicity of the algorithms belonging to these classes creates a limited variability among the binaries.
  The {\sf M} class is also one of the classes where the Gemini embeddings are performing better, this is probably due to the fact that one of the manual features used by Gemini is the number of arithmetic assembly instructions inside a code block of the CFG.
By analyzing the output of the classifier we find out that the most common error, a mistake common to both Gemini case and SAFE, is the confusion between encryption and sorting algorithms. A possible explanation for this behaviour is that simple
 encryption algorithms, such as RC5, share many similarities with sorting algorithms (e.g., nested loops on an array).  

 Finally, we can see that, in all cases, the embeddings computed with our architecture outperform the ones computed with Gemini; the improvement range is between $10\%$ and $2\%$. The average improvement, weighted on the cardinality of each class, is around $6\%$.

\paragraph{Qualitative Analysis of the Embeddings}
We performed a qualitative analysis of the embeddings produced with SAFE. Our aim is to understand how the network captures the information on the inner semantics of the binary functions, and how it represent such information in the vector space.
\begin{figure*}[t!]
\center

\includegraphics[width=.9\textwidth]{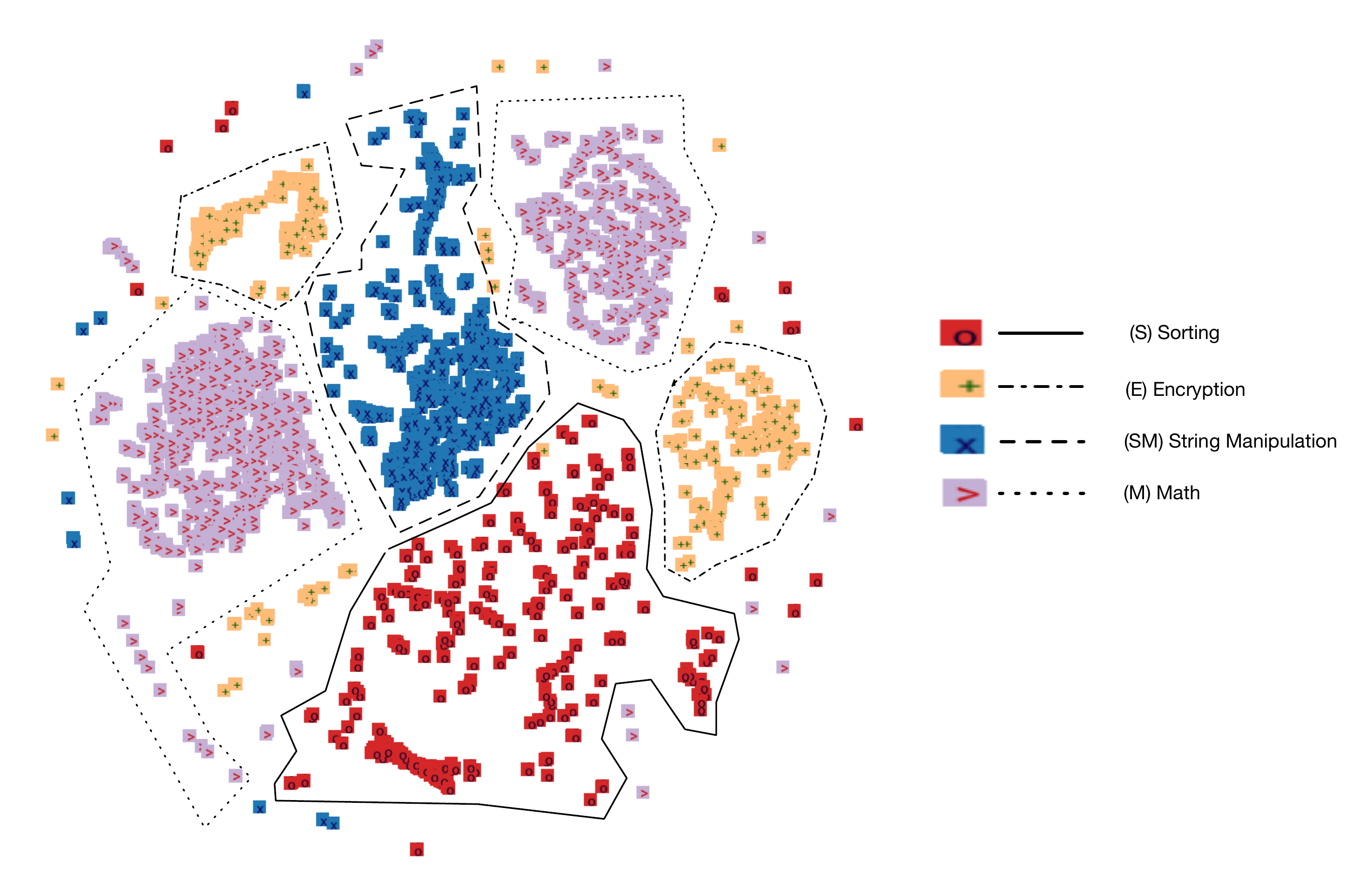}
\caption{2-dimensional visualization of the embedding vectors for all binary functions in \simDataset. The four different categories of algorithms (Encryption, Sorting, Math and String Manipulation) are represented with different symbols and colors. \label{sf:2}}
\end{figure*}

To this end we computed the embeddings for all functions in \simDataset. In 
Figure \ref{sf:2} we report the two-dimensional projection of the $100$-dimensional vector space where binary functions embeddings lie, obtained using the \emph{t-SNE} \footnote{We used the TensorBoard implementation of \emph{t-SNE}} visualisation technique \cite{maaten2008visualizing}.
From Figure \ref{sf:2} is possible to observe a quite clear separation between the different classes of algorithms considered. We believe this behaviour is really interesting and it further confirms our quantitative experiments on semantic classification. 

\paragraph{Real use case of Task 4 - Detecting encryption functions in Windows Malwares}
We decided to test the semantic classifcation on a real use case scenario.
We trained a new SVM classifier using the semantic dataset with only two classes, encryption and non-encryption. We then used this classifier on malwares. 
 We analyzed two samples of window malwares found in famous malware repositories: the {\em TeslaCrypt} and {\em Vipasana} ransomwares.
We disassembled the malwares with radare2, we included in the caller the code of the callee functions up to depth 2.
We processed the disassembled functions with our classifier, and we selected only the functions that are flagged as encryption with a probability score greater than $96\%$.
Finally, we manually analyzed the malware samples to assess the quality of the selected functions.
\begin{itemize}
\item {\em TeslaCrypt} \footnote{The sample is available at url \url{https://github.com/ytisf/theZoo/tree/master/malwares/Binaries/Ransomware.TeslaCrypt}. The variant analyzed is the one with hash {\it 3372c1eda...} }: on a total of 658 functions, the classifier flags  the ones at addresses {\sf 0x41e900}, {\sf 0x420ec0}, {\sf 0x4210a0}, {\sf 0x4212c0}, {\sf 0x421665}, {\sf 0x421900}, {\sf 0x4219c0}.  We confirmed that these are either encryption (or decryption) functions or helper functions directly called by the main encryption procedures. 

\item {\em Vipasana}  \footnote{The sample is available at url \url{https://github.com/ytisf/theZoo/tree/master/malwares/Binaries/Ransomware.Vipasana}. The variant analyzed is the one with hash {\it 0442cfabb...4b6ab} }: on a total of 1254 functions, the classifier flags the ones at addresses {\sf 0x406da0}, {\sf 0x414a58}, {\sf 0x415240}. We confirmed that two of these are either encryption (or decryption) functions or helper functions directly called by the main encryption procedures. The false positive is {\sf 0x406da0}. 

\end{itemize} 

As final remark, we want to stress that these malwares are for windows and they are 32-bit binaries, while we trained our entire system on ELF executables for AMD64. This shows that our model is able to generate good embeddings also for cases that are largely different from the ones seen during training.

%% file: sec6_conclusion.tex

\section{Speed considerations.}\label{sec;speed}
As reported in the introduction, one of the advantages of SAFE that it ditches the use of CFGs. From our tests on radare2 disassembling a function is $10$ times faster than computing its CFG. Once functions are disassembled an Nvidia K80 running our model computes the embeddings of $1000$ functions in around $1$ second.

More precisely, we run our tests on a virtual machine hosted on Google cloud platform. The machine has 8 core Intel Sandy Bridge, 30gb of ram, an Nvidia K80 and SSD hard-drive.  
We disassembled all object files in postegres 9.6 compiled with gcc-6 for all optimizations.  During the disassembling we assume to know the starting address of a function, see \cite{functionstart} for a paper using neural networks to find functions in a binary. 

The time needed to disassemble and pre-process 3432 binaries is 235 seconds, the time needed to compute the embeddings of the resulting 32592 functions is $33.3$ seconds. The end-to-end time to compute embeddings for all functions in postgres starting from binary files is less than 5 minutes. 
We repeated the same test with openssl 1.1.1 compiled with gcc-5 for all optimizations. The end-to-end time to compute the embeddings for all functions in openssl is less than 4 minutes.

Gemini is up to 10 times slower, it needs 43 minutes for postgres and 26 minutes for openssl.

\section{Conclusions and future works}
\label{sec:conc}

In this paper we introduced SAFE an architecture for computing embeddings of functions in the cross-platform case that does not use debug symbols. 
Our architecture does not need the CFG, and this leads to a considerable speed advantage. SAFE creates thousand embeddings per second on a mid-CTOS GPU. Even when we factor the disassembling time our end-to-end system (from binary file to function embedding), processes more than 100 functions per second. 
This considerable speed comes with a significant increase of predictive performances with respect to the state of the art. Summing up, SAFE is both faster and more precise than previous solutions.

Finally, we think that our experiments on semantic detection are really interesting, and they pave the way to more complex and refined analysis, with the final purpose of building binary classifiers that rival with the classifiers today available for image recognition. 

\paragraph{Future Works}
There are several immediate lines of improvement that we plan to investigate in the immediate future. The first one is to retrain our i2v model to make use of libc call symbols. This will allow us to quantify the impact of such information on embedding quality. We believe that symbols could lead to a further increase of performance, at the cost of assuming more information and the integrity of the binary that we are analyzing.

 The use of libc symbols would enable a more fine grained semantic classification: as example we could be able to distinguish a function that is sending encrypted content on a socket from a function that is writing encrypted content on a file.
 Summarizing, the field of applied machine learning for binary analysis is still in its infancy and there are several opportunities for future works.\\

\medskip

{\bf Acknowledgments.} 
The authors would like to thank Google for providing 
free access to its cloud computing platform through the Education Program. Moreover, the authors would like to thank NVIDIA for partially supporting this work through the donation of a GPGPU card used during prototype development. Finally, the authors would like to thank  Davide
Italiano for the insightful discussions.
This work is supported  by a grant of the Italian Presidency of the Council of Ministers and by the CINI (Consorzio Interuniversitario Nazionale Informatica) National Laboratory of Cyber Security.